# Highlights from RHIC Spin Physics Program

Ming X. Liu

*P-25, MS H846, Los Alamos National Laboratory, Los Alamos, NM 87545*

**Abstract.** The Relativistic Heavy Ion Collider (RHIC) at Brookhaven National Laboratory delivers the world's highest energy polarized proton-proton collisions at a center of mass energy up to 500 GeV and provides a unique opportunity to study the quark and gluon spin structure of the proton and QCD dynamics at high energy scale. RHIC has produced many exiting physics results in recent years. The latest data from RHIC significantly constrain the gluon spin contribution to the proton spin, and the parity violating single spin asymmetry are observed for the first time in W production by both the PHENIX and STAR collaborations. In this report, I present the latest results from the PHENIX and STAR experiments, followed by a brief discussion of the future prospects of transverse physics, particularly on the importance of the unique measurements of Drell-Yan single spin asymmetry.

## INTRODUCTION

Spin plays a key role in the determination of the properties of fundamental particles and their interactions. However, two outstanding spin puzzles have challenged our understanding of QCD over 30 years: (1) the challenge of "too small", i.e., the nucleon spin puzzle. It was believed that the spin of proton was carried mostly by the quarks that make up the proton. However, experiments in the 1980's led to the startling discovery that quark-spin contribute very little to the proton spin, setting off the "proton spin crisis". Where the rest of the proton spin coming from remains a major challenge to our understanding of the nucleon structure. Currently, there are several candidates that could make up the difference, the gluons' spin and quark and gluon's orbital angular momentum. Their contributions need to be determined by experimental measurements; (2) the challenge of "too big", i.e., the unexpectedly large left-right transverse single spin asymmetries (TSSA, $A_N$) in transversely polarized high-energy $p^\uparrow + p$ (and $l + p^\uparrow$) collisions. Prior to the early Fermilab E704 experimental measurements, it was believed based on collinear pQCD arguments that transverse single spin asymmetry should be very small in high-energy p+p collisions, $A_N \sim \frac{m_q}{p_T} \sim O(10^{-4})$, while the experimental data showed asymmetry order of 10%. Such large TSSAs are also observed at RHIC energies. Spin measurements have historically produced surprising results and are a stringent test to theories as spin is an intrinsically relativistic and quantum mechanical aspect of particle interactions.

## Highlights from the Longitudinal Spin Physics Program

Measurements of the spin structure of the nucleon are a critical step in our understanding of strong interaction and nucleon structure. High energy polarized proton collisions at a center of mass energy up to $\sqrt{s}$ =500 GeV at RHIC provide a unique way to probe the proton structure. At RHIC energies, the NLO pQCD calculations have been applied successfully to describe the unpolarized cross section of hadrons and jet productions and also have been used to extract the polarized parton (quark and gluon) distribution functions.

## Gluon Spin Contribution to the Proton Spin

Since the discovery that quark's spin only contributes ~30% of the total proton spin, the gluon polarization inside the proton has been one of the major focus of experimental and theoretical investigations. Unlike the polarized DIS process, where gluons do not couple directly to the virtual phones, polarized p+p interactions provide direct access to the polarized quark and gluon distribution functions via the hard scattering processes at the leading order. The non-perturbative but universal polarized parton distribution functions $\Delta f_a(x,Q^2)$ can be accessed via the measurements of double spin asymmetries,

$$A_{LL} \equiv \frac{d\Delta\sigma}{d\sigma} \equiv \frac{d\sigma^{++} - d\sigma^{+-}}{d\sigma^{++} + d\sigma^{+-}}$$

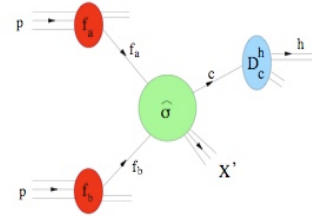

with spin dependent particle $h$ (or jet) production cross section given by,

$$d\Delta\sigma^h \sim \sum_{a,b} \int dx_a \int dx_b \Delta f_a(x_a,Q^2) \Delta f_b(x_b,Q^2) \times d\Delta\hat{\sigma}^{a+b \rightarrow c+X} \times D^{c \rightarrow h}$$

where the sum runs over all initial state partons $a$ and $b$, with $d\Delta\hat{\sigma}^{a+b \rightarrow c+X}$ the corresponding partonic cross sections and $D^{c \rightarrow h}$ the fragmentation functions.

Recently, the PHENIX and STAR experiments have reported precision measurements of longitudinal double spin asymmetry $A_{LL}$ in inclusive high $p_T$ neutral pion production and inclusive high pT jet production in polarized p+p collisions at √s = 200 GeV, respectively[1,2], see Figure 1.

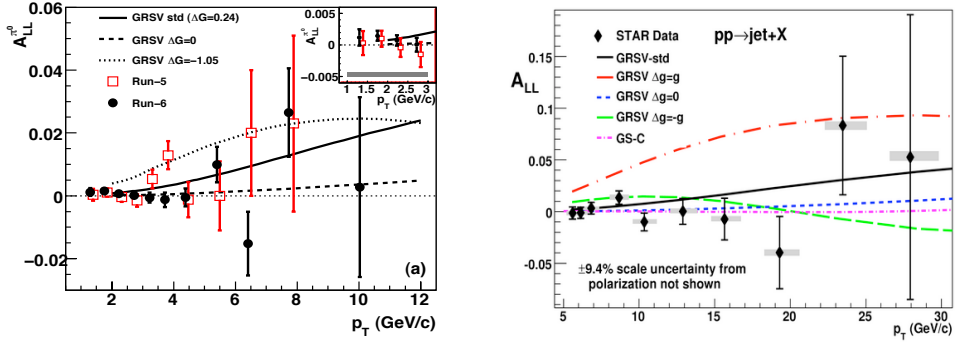

**FIGURE 1.** Left: Asymmetry $A_{LL}$ in $\pi^0$ production as a function of $p_T$, from PHENIX. Right: Inclusive Jet $A_{LL}$ as function of jet $p_T$, from STAR.

The results from PHENIX and STAR have put a strong constrain on the gluon spin contribution to the proton total spin. The details depend on the polarized parton distribution models used for the fit. In the GRSV framework[3], the PHENIX experiment has put a limit on gluon spin contribution, $-0.7 < \Delta G^{[0.02, 0.3]} < 0.5$, at $\Delta\chi^2 = 9$. Much improved measurements are expected from future high luminosity runs in coming years.

## Parity Violating Single Spin Asymmetry $A_L$ in $W^{+/-}$ production

The parity violating longitudinal single spin asymmetry $A_L$ in $W^{+/-}$ production in polarized p+p collisions is sensitive to the flavor dependent $\Delta q$ and $\Delta \bar{q}$ distributions as W bosons only couple to the left-handed fermions and right-handed anti-fermions,

$$u_L + \bar{d}_R \rightarrow W^+$$
$$d_L + \bar{u}_R \rightarrow W^-$$

At leading order, the parity violating asymmetry is given by,

$$A_L(W^+) = \frac{\sigma^+ - \sigma^-}{\sigma^+ + \sigma^-} = \frac{\Delta\bar{d}(x_1)u(x_2) - \Delta u(x_1)\bar{d}(x_2)}{\bar{d}(x_1)u(x_2) + u(x_1)\bar{d}(x_2)}$$

$$A_L(W^-) = \frac{\sigma^+ - \sigma^-}{\sigma^+ + \sigma^-} = \frac{\Delta\bar{u}(x_1)d(x_2) - \Delta d(x_1)\bar{u}(x_2)}{\bar{u}(x_1)d(x_2) + d(x_1)\bar{u}(x_2)}$$

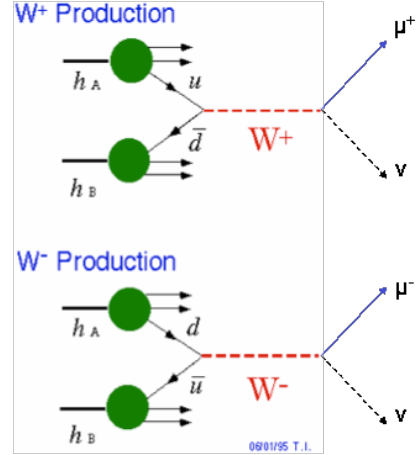

In 2010, the PHENIX and STAR collaborations reported the first observations of W boson production and significant non-zero parity violating single spin asymmetry from the 2009 run data[4,5]. Figure 2 shows the results from the STAR experiment, measurements are consistent with pQCD expectations. Future high statistic data will provide first clean measurements of flavor-identified sea-quark polarization that is almost impossible in DIS and further test pQCD predictions in a much wider kinematic range.

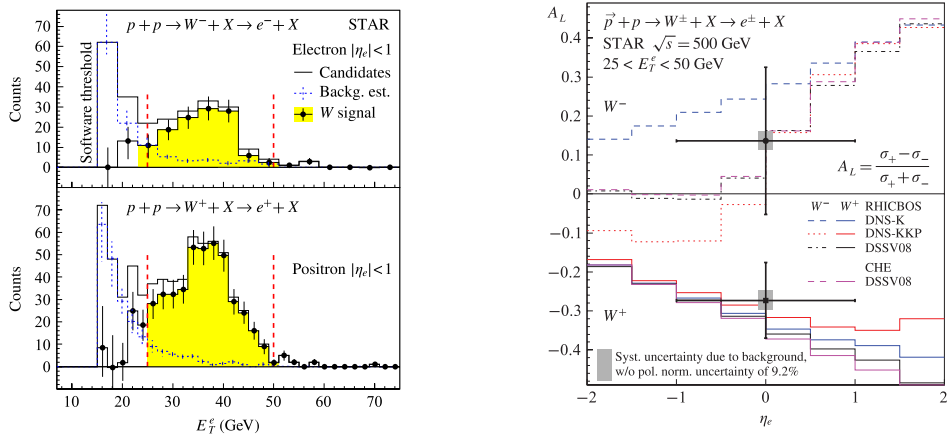

**FIGURE 2.** Left: high $p_T$ electrons from W decays measured by the STAR experiment in the central rapidity; Right: high $p_T$ electrons parity violationg single spin asymmetries for W events.

## Highlights from the Transverse Spin Physics Program

Since the beginning of RHIC-Spin, interesting results on transverse single spin asymmetries and cross sections have been measured by the RHIC spin experiments (BRAHMS, PHENIX and STAR) in a very wide kinematic range in polarized p+p collisions. All three spin experiments have observed and confirmed significant non-zero TSSAs in inclusive single particle productions in the forward rapidities. In the middle and backward rapidities, TSSAs are consistent with zero within the experimental uncertainty.

The STAR collaboration has measured $\pi^0$ TSSA as a function of $x_F$ and $p_T$, up to $p_T \sim 4.5$GeV from the latest transverse run in 2008. The PHENIX experiment has measured TSSAs in pion and inclusive charged hadron production as a function of $x_F$ and $p_T$ in the central and forward rapidities and confirmed previous results from other experiments. PHENIX also carried out the first measurements of TSSAs in open heavy quark and J/Psi production at $\sqrt{s}$=200 GeV, although statistically limited, the preliminary results indicate possible interesting new physics for transverse spin program in the future high luminosity runs.

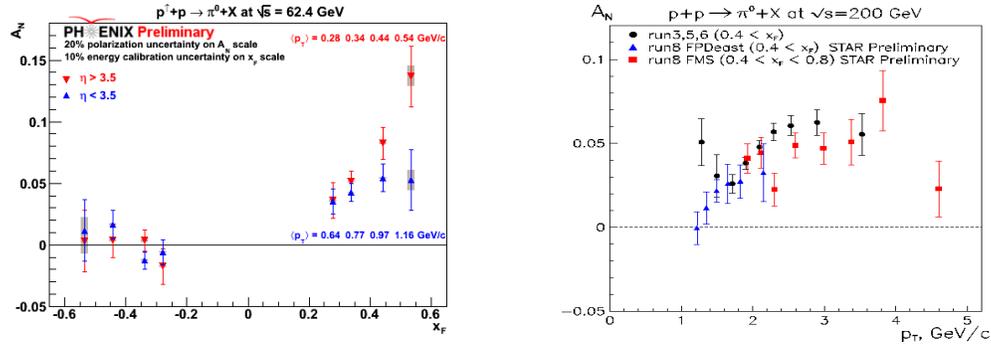

**FIGURE 3.** Left: TSSA $A_N$ in $\pi^0$ production as a function of $x_F$ at the center of mass energy $\sqrt{s}$=62 GeV, measured by the PHENIX experiment; Right: $A_N$ in $\pi^0$ production as a function of $p_T$ at the center of mass energy $\sqrt{s}$=200 GeV, from the STAR experiment.

With upcoming detector upgrades for both the PHENIX and STAR experiments and expected much improved accelerator performance, RHIC-Spin program will provide a great opportunity for new studies of polarized nucleon structure and the QCD dynamics. The future Drell-Yan $A_N$ measurement is particularly important at RHIC[6]. There is a fundamental prediction of QCD that $A_N$ of Drell-Yan should have an opposite sign compared to the $A_N(\pi^+)$ observed in the polarized DIS experiments, see Figure 4. Its verification or disproval will be an important milestone in our understanding of QCD.

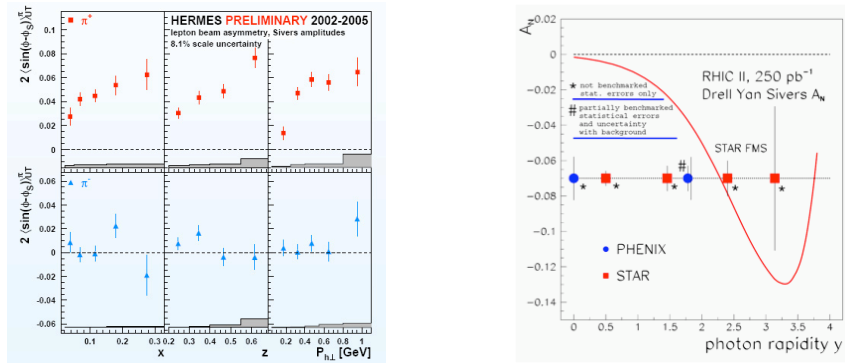

**FIGURE 4.** Left: Charged pion Sivers asymmetries measured by the HERMES experiment. Right: Expected Drell-Yan TSSA $A_N$ in p+p at RHIC (based on HERMES fit), the sign of the asymmetry should be opposite to the one observed in $\pi^+$ production in the polarized DIS process.

## ACKNOWLEDGMENTS

We thank the BNL Collider-Accelerator Department for developing the unique technologies enabling these measurements, and colleagues from PHENIX and STAR collaborations for helpful discussions. We acknowledge support from the Office of Nuclear Physics in DOE Office of Science.